\documentclass[pra,twocolumn,showpacs]{revtex4}
\usepackage{graphicx}
\usepackage{amsmath}
\usepackage[abs]{overpic}
\usepackage{rotating}
\usepackage{hyperref}
\newcommand{\ket}[1]{|#1\rangle}
\newcommand{\bra}[1]{\langle #1|}

\begin{document}
\title{Geometric Phase in Entangled  Bipartite Systems}
\author{H. T. Cui}\email{cuiht@student.dlut.edu.cn}
\author{L. C. Wang, X. X. Yi} \homepage{http://qpi.dlut.edu.cn}
\affiliation{Department of Physics, Dalian University of Technology,
Dalian 116024, China}

\date{\today}
\begin{abstract}
The geometric phase (GP) for bipartite systems in transverse
external magnetic fields is investigated in this paper. Two
different situations have been studied. We first consider two
non-interacting particles. The results show that because of
entanglement, the geometric phase is very different from that of the
non-entangled case. When the initial state is a Werner state, the
geometric phase is, in general, zero and moreover the singularity of
the geometric phase may appear with a proper evolution time. We next
study the geometric phase when intra-couplings appear and choose
Werner states as the initial states to entail this discussion. The
results show that unlike our first case, the absolute value of the
GP is not zero, and attains its maximum when the rescaled coupling
constant $J$ is less than 1. The effect of inhomogeneity of the
magnetic field is also discussed.
\end{abstract}
\maketitle

\section{introduction}
A system can retain the information of its motion when it undergoes
a cyclic evolution, in the form of a geometric phase (GP), which was
first put forward by Pancharatnam in optics \cite{p}  and later
studied explicitly  by Berry in a general quantal
system\cite{berry}. Great progress has been made in this novel
region \cite{shapere}; The original adiabatic condition in Berry's
work has been removed by Aharonov  and Anandan \cite{aharonov}, and
Samuel and Bhandari have generalized the geometric phase by
extending  to noncyclic evolution and sequential measurements
\cite{samuel}. At the same time a kinematic approach to the theory
of the geometric phase has also been developed by Mukunda and Simon
\cite{mukunda1}. Recently  the generalization to mixed states was
conducted first by Uhlmann in the mathematical context of
purification \cite{uhlmann}, and then by  Sj\"{o}qvist  {\it et al.}
based on the Mach-Zender interferometer \cite{erik}. Consequently
the mixed-state geometric phase has been experimentally verified
using both NMR interferometry \cite{du} and single photon
interferometry \cite{ericsson}. Recently the geometric phase for
mixed state was put forward by   Singh \textit{et al.} for the case
of degenerate density operators \cite{singh} and a general formula
for the parallel transporting was also provided. Despite the great
progress in this field, Bhandari recently raised the criticism that
the definition for geometric phase in mixed state fails when the
interference fringes disappear \cite{bhandari}. This  can be
explained as the disappearance (appearance)  of the geometric phase
(off-diagonal geometric phase). The definition of off-diagonal
geometric phase(OP) was first given by Manini \textit{et al.} for
pure states in adiabatic evolutions\cite{manini}, and then was
generalized to non-adiabatic situation\cite{mukunda2} and in
mixed-states\cite{filipp}. Moreover the off-diagonal geometric phase
was studied in  degenerate case \cite{tong1}and in bipartite
systems\cite{yi1}. Recently the effect of  entanglement on the
off-diagonal geometric phase has also discussed \cite{cui}.

The quantum computation scheme for the geometric phase has  been
proposed based on the Abelian \cite{jones} or non-Abelian geometric
phase  \cite{zanardi}, in which geometric phase has been shown to be
intrinsic against faults in the presence of some kind of external
noise due to the geometric nature of Berry phase. Consequently the
quantum gates based on geometric phase have also been proposed in
different systems \cite{zheng}, where the interactions play an
important role for the realization of some specific operations.

Bipartite system is of great importance in quantum computation, such
as the transfer  of quantum information, the construction of
entanglement as well as the realizations of logic operations.
Furthermore, it was found that GPs may be used to design quantum
logic gates. These facts together give rise to a question what are
the geometric phase and its motion in bipartite systems. Recently
some papers have addressed this issue \cite{erik2, yi2}, where the
discussions respectively focus on the entanglement dependence of the
geometric phase for subsystem and the coupling effect on the GP for
subsystem under adiabatical evolution. However, another question
remains open; how  the entanglement or interaction affects the
geometric phase for the whole system. This consideration is not
trivial since any  quantum information procession cannot implemented
 by only one qubit. Moreover some interesting results may be
found with the increment of the size of the system. In this paper,
focusing on the entanglement and inter-subsystem couplings, we
present explicit discussion on the geometric phases in bipartite
systems.

For this purpose, our discussion is divided into several sections.
In Sec. II, we describe the model to be discussed in this paper and
some formulas are also present for the  calculation of  the
geometric phase. Then  in Sec. III we study the geometric phase for
two noninteracting particles and  some interesting results can be
found in this section. The intra-subsystem coupling effect on the
geometric phase is studied in Sec. IV. Finally, conclusions are
given at the end of this paper.

\section{model}
In order to highlight the effect of the entanglement, we choose the
initial state as
\begin{equation}\label{rho}
\rho(0)=\frac{1-r}{4}I+r\ket{\Phi}\bra{\Phi},
\end{equation}
where $r\in(0,1]$ determines the mixing of this state and $I$ is the
unit matrix in the  $2\times2$ Hilbert space. Notice  that for $r=0$
Eq. \eqref{rho} is the unit matrix and its geometric phase is
trivial (always be zero), our discussion excludes this case. The
state $\ket{\Phi}$ may be either of the following states,
\begin{eqnarray}\label{}
\ket{\varphi}&=\sin\theta\ket{11}_{1,2}
                      +\cos\theta\ket{00}_{1,2} \nonumber\\
\ket{\psi}&=\sin\theta\ket{10}_{1,2}
                   +\cos\theta\ket{01}_{1,2}
\end{eqnarray}
where $\theta$ determines the degree of entanglement and
$\ket{1(0)}_{i}(i=1,2)$ is the eigenstate of  Pauli operator
$\sigma_{z}$. One should note that when $\theta=(3)\pi/4$, the
equations above are Bell states and Eq.(\ref{rho}) are the Werner
states \cite{werner}, which plays an important role in quantum
information processings, especially in quantum communication via
noisy channels \cite{bennett1} and quantum distillation
\cite{bennett2}. In general the mixed state Eq. (\ref{rho}), which
first was introduced by Wootters \cite{wootters}, may be entangled;
that can be judged  by the Peres-Horodecki condition \cite{peres}.
Eq. (\ref{rho}) includes all possible cases, such as pure or mixed
state and maximal or non-maximal entangled states. One should note
that Eq. (\ref{rho}) is triplet-degenerate for $r\neq1$.

We should point out that the initial state Eq. \eqref{rho} is not a
trivial generalizing from the pure case. The first term in Eq.
\eqref{rho} can be regarded as the noise and the mixing coefficient
$r$ properly describe the intensity of noise. Recently the
one-to-one correspondence between $r$ of Werner state and the
temperature $T$ of the one-dimensional Heisenberg two-spin chain
with a magnetic field $B$ along the $z$-axis, has been established
\cite{batle}. This connection gives us the strong physical support
for the initial state Eq. \eqref{rho}.

We choose the system composed of two spin-1/2 particles undergoing
spin procession in an external time independent magnetic field in
the $z$ direction. Then the Hamiltonian is
\begin{equation}\label{h}
H=H_{0}+H_{I},
\end{equation}
in which the free Hamiltonian $H_0$ and the XX interaction  $H_I$ are respectively
\begin{eqnarray}
H_{0}&=&\omega_{1}S_1^z+\omega_{2}S_2^z\nonumber \\
H_{I}&=&g(S^{\dag}_{1} S^{-}_2 + S^{-}_1 S^{\dag}_2),
\end{eqnarray}
where $g>0$ is the antiferromagnetism coupling constant and
$S_i^z(i=1,2)$ is the $z$ component of spin operator respectively.
$S^{\pm}_{i}=S_{i}^x+iS_{i}^y(i=1,2)$ are the raising and decreasing
operators of the $z$ component of spin-1/2. Actually the Hamiltonian
Eq. (\ref{h}) is  two-qubit XX model, which is of  fundamental
importance to  understand the relation between the entanglement and
quantum correlation  in interacting many-body systems. In general we
suppose that $\omega_1$ may not be equal to $\omega_2$ because of
the inhomogeneity of the external magnetic field. One will find in
the following discussion that the inhomogeneity of the external
magnetic field has nontrivial effect on GP.

Some formulas should be addressed for the  calculation of  GP in
this model. Recently GP for the mixed state has been discussed by
Sj\"oqvist \textit{et al.}in \cite{erik}, based on the Mach-Zender
interferometer and a formula was provided for calculation of GP for
a mixed state,
\begin{equation}\label{}
\gamma_{g}=\textrm{Arg}\textrm{Tr}[U^{\|}(t)\rho(0)],
\end{equation}
in which $U^{\|}(t)$ was defined as the parallel transportation
operator. Based on this work, Singh, \textit{et al.,} \cite{singh}
studied GP for non-degenerate and degenerate mixed states and
provided a general method for the construction of $U^{\|}(t)$  by
imposing
\begin{equation}\label{}
U^{\|}(t)=U(t)V(t)
\end{equation}
in which $U(t)$ is the unitary evolution operator and is equal to $
e^{- iHt}$ in our model and $V(t)$ is a blocked matrix, of which the
elements are determined by
\begin{eqnarray}
V_{\mu\nu}&=&\bra{\mu}e^{it\sum_{\mu',\nu'} \bra{\mu '}H\ket{\nu
'}\ket{\mu
'}\bra{\nu '}}\ket{\nu},\nonumber\\
&&\ket{\mu}, \ket{\nu} , \ket{\mu '}, \ket{\nu '}\in\{degenerate\ \  subspace\} \nonumber\\
V_{kk} &=& e^{i \bra{k}H\ket{k}t}, \ket{k}\in\{the\  remaining \
space\},
\end{eqnarray}
where $\ket{\mu}, \ket{\nu}, \ket{\mu '}, \ket{\nu '}, \ket{k}$  are
the eigenstates of $\rho(0)$, and the interference terms between the
degenerate space and the other space are set to be zero in order to
keep the parallel transporting in the degenerate space. One notes
that for different mixed states one has different $U^{\|}(t)$. Based
on these formulas, one can calculate GP. In the following
calculations, we label the  initial states as
\begin{eqnarray}
\rho_1&=&\frac{1-r}{4}I_{4}+r\ket{\varphi}\bra{\varphi} \nonumber \\
\rho_2&=&\frac{1-r}{4}I_{4}+r\ket{\psi}\bra{\psi}.
\end{eqnarray}
and the geometric phases for these states are respectively
calculated in the following parts.

\section{ $g=0$ case}
We first study the geometric phase without interaction. It should
point out that this situation still has interest  in quantum
information, such as the quantum teleportation and communication in
which the nonlocal correlation, i. e. the entanglement between two
space-liked particles plays a crucial role. The geometric phases for
$\rho_{i}$(i=1,2)in this case can be easily obtained from the
formulas in the former section,
\begin{widetext}
\begin{eqnarray}\label{g1}
\gamma_{g1}&=&\arctan\frac{-r(\sin[\cos2\theta\frac{\omega_1+\omega_2}{2}
t]\cos\frac{\omega_1+\omega_2}{2}t - \cos2\theta
\cos[\cos2\theta\frac{\omega_1+\omega_2}{2}
t]\sin\frac{\omega_1+\omega_2}{2}t)}
{\frac{1-r}{2}+\frac{1+r}{2}(\cos[\cos2\theta\frac{\omega_1+\omega_2}{2}
t]\cos\frac{\omega_1+\omega_2}{2}t + \cos2\theta
\sin[\cos2\theta\frac{\omega_1+\omega_2}{2}
t]\sin\frac{\omega_1+\omega_2}{2}t)}\nonumber\\
\gamma_{g2}&=&\arctan\frac{-r(\sin[\cos2\theta\frac{\omega_1-\omega_2}{2}
t]\cos\frac{\omega_1-\omega_2}{2}t - \cos2\theta
\cos[\cos2\theta\frac{\omega_1-\omega_2}{2}
t]\sin\frac{\omega_1-\omega_2}{2}t)}
{\frac{1-r}{2}+\frac{1+r}{2}(\cos[\cos2\theta\frac{\omega_1-\omega_2}{2}
t]\cos\frac{\omega_1-\omega_2}{2}t + \cos2\theta
\sin[\cos2\theta\frac{\omega_1-\omega_2}{2}
t]\sin\frac{\omega_1-\omega_2}{2}t)},
\end{eqnarray}
\end{widetext}

\begin{figure}[tbp]
\begin{center}
\begin{overpic}[width=4cm]{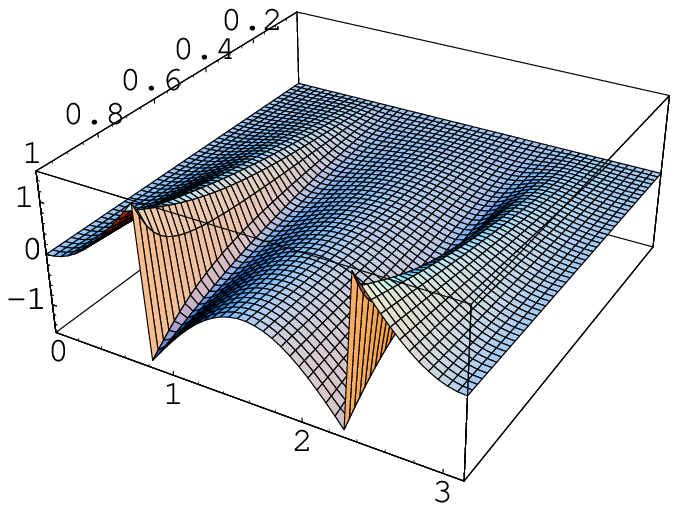}
\put(40,10){$\theta$} \put(20,80){$r$} \put(-8,60){$\gamma_{g1}$}
\put(0,75){(a)}
\end{overpic}
\begin{overpic}[width=4cm]{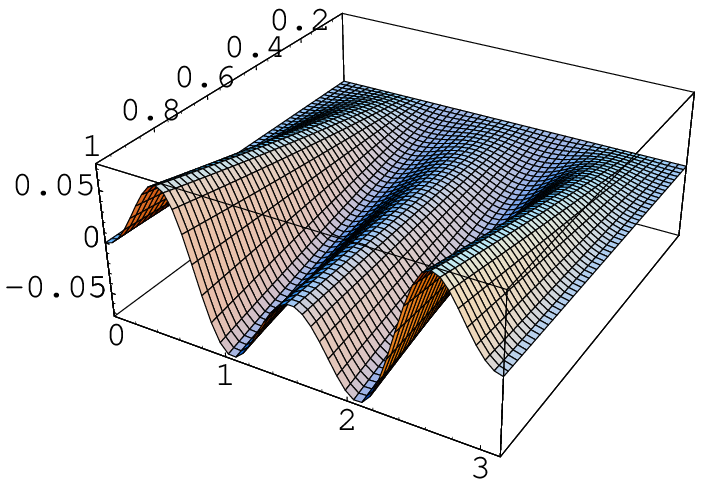}
\put(40,10){$\theta$} \put(20,70){$r$} \put(0,60){$\gamma_{g2}$}
\put(0,75){(b)}
\end{overpic}
\caption{ \label{1} The geometric phases $\gamma_{gj}(j=1,2)$[Arc]
versus  $r, \theta$. For $\gamma_{g1}$ (a), we have chosen
$\omega_{1}= \omega_{2}$ and $\omega_{1}t=\pi/2$; whereas for
$\gamma_{g2}$ (b), $\omega_{1}= 2 \omega_{2}$ and
$\omega_{1}t=\pi$.}
\end{center}
\end{figure}

\begin{figure}[t]
\begin{center}
\begin{overpic}[width=4cm]{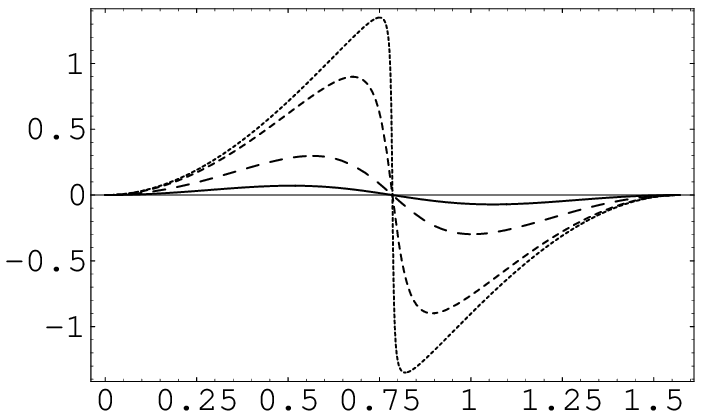}
\put(-5,32){\begin{rotate}{90} $\gamma_{g1}$
\end{rotate}}
\put(55,-5){$\theta$} \put(15, 55){(a)}
\end{overpic}
\begin{overpic}[width=4cm]{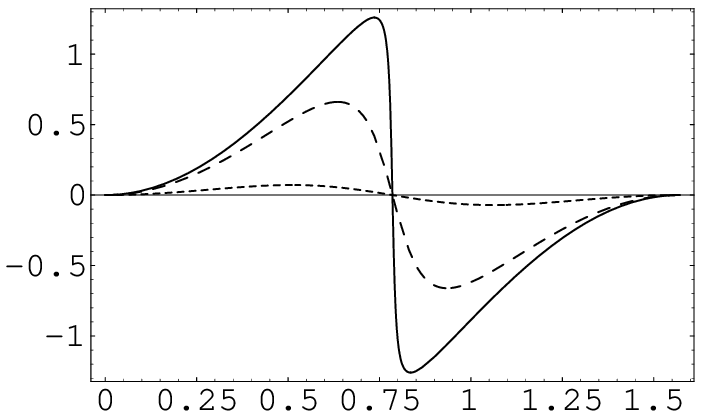}
\put(3,32){\begin{rotate}{90}$\gamma_{g2}$\end{rotate}}
\put(55,-5){$\theta$} \put(15, 55){(b)}
\end{overpic}
\\[0.5cm]
\begin{overpic}[width=4cm]{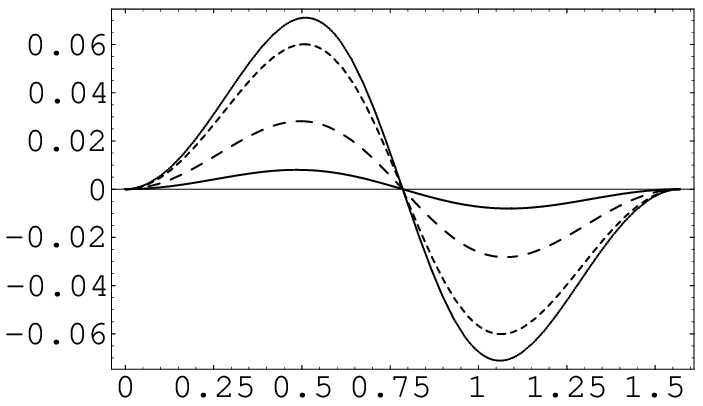}
\put(-5,32){\begin{rotate}{90} $\gamma_{g1}$
\end{rotate}}
\put(55,-5){$\theta$} \put(18, 53){(c)}
\end{overpic}
\begin{overpic}[width=4cm]{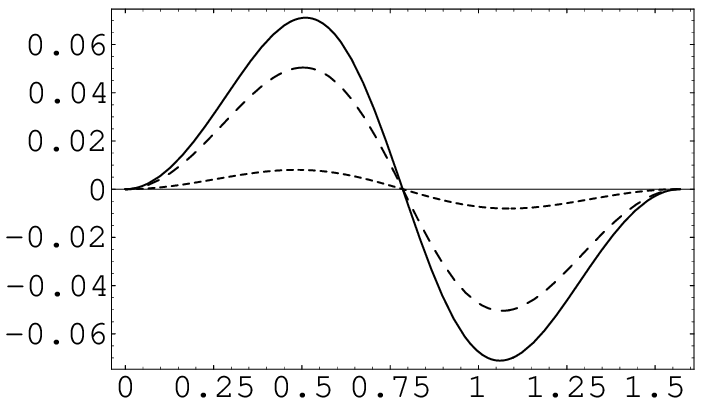}
\put(3,32){\begin{rotate}{90} $\gamma_{g2}$
\end{rotate}}
\put(55,-5){$\theta$} \put(18, 53){(d)}
\end{overpic}
\caption{\label{2} The geometric phase [Arc] vs the inhomogeneity of
magnetic fiedl $n=\omega_2/\omega_1$ with $r=1$. Since the figures
is symmetrical with the $\theta=\pi/2$, we only draw for
$\theta\in[0, \pi/2]$. (a) corresponds to $\rho_{1}$ and the other
parameters are the same as that of Fig. \ref{1}(a). The dotted,
dashed, longer-dashed and solid lines correspond respectively to
$n=0.99, 0.9, 0.5, 0$. When $n=1$, the geometric phase is not
discontinued at the point $\theta=\pi/4$. (b) is for $\rho_{2}$ and
the other parameters are the same as that of Fig. \ref{1}(b). The
dotted, dashed and solid lines correspond to $n=0.5, 0.1, 0.01$.
When $n=0$, the geometric phase is not discontinued in the point
$\theta=\pi/4$. (c) and (d) demonstrate the case that there is no
the singularity ( We have chosen $\omega_1t=\pi/4$ for $\rho_1$ and
$\omega_1t=\pi/2$ for $\rho_2$ respectively).}
\end{center}
\end{figure}

It is interesting to note that when the initial states are Werner
states, the geometric phase is  zero when the denominators in
Eqs.\eqref{g1}  are not vanishing simultaneously. Furthermore when
$\omega_1=\omega_2$, the geometric phase for $\rho_2$ is zero since
$\rho_2$ is commutative with the Hamiltonian and it cannot pick up
any geometric phase. A detailed demonstration for
$\gamma_{gj}(j=1,2)$ with the parameters $r, \theta$ is shown in
Fig. \ref{1}. From the figures it is obvious that the absolute
values of $\gamma_{g1(2)}$ attain the maximum when $\theta\neq\pi/4,
3\pi/4$ and because of the mixing, scaled by $r$, the absolute value
of geometric phase is compressed and tends to be zero with
$r\rightarrow 0$. We also note that a singularity about
$\gamma_{g1}$ appears when $\theta=(3)\pi/4$ and $r=1$, as displayed
in Fig. \ref{1}(a). At this point, the numerator and the dominator
in the expression of $\gamma_{g1}$ in Eq. \eqref{g1} are
simultaneously zero and the geometric phase is undefined.  One has
to calculate the so-called off-diagonal geometric phase to retain
the information of the evolution \cite{manini}. Furthermore, our
calculation shows that the singularity depends not only on  the
entanglement, but on how the system evolves.

With the consideration of the two noninteracting particles, it is of
interest to discuss the effect of the inhomogeneity of the external
fields. The results have been illustrated in Fig. \ref{2}. It
clearly shows that  when there is a singular point, for $\rho_{1}$,
the absolute value of the geometric phase with $n\rightarrow1$
attains the maximum closed to this point. Whereas,  for $\rho_{2}$,
this happens for  $n\rightarrow0$. We should emphasize that this
phenomenon appears only when there exists the singularity, which is
induced by the entanglement of the initial state. If there is no
possibility of the appearance of the singularity, our calculation
shows that the points where  the absolute value of the geometric
phase attains the maximum is independent of the inhomogeneity of the
external field,just shown as Fig.\ref{2}(c)and (d) and always is
zero for the Werner state. Physically it is thus possible for this
phenomenon to act as the signature of the singularity of the
geometric phase, which is usually originated from the degenerace
\cite{manini}.

Another interesting consideration is that one supposes
$\omega_{2}=0$, which corresponds to case that particle A is
processing with frequency $\omega_{1}$ while particle B keep
dynamically fixed. In this case, the affect from the dynamics of the
second particle is excluded and one can more clearly check the
effect of the entanglement on the geometric phase. The geometric
phase for this special situation can be obtained by setting
$\omega_2=0$ in Eq. \eqref{g1}. One can  find that because of the
entanglement, the geometric phase for particle 1 is zero too for
Werner state. and it is also possible for the singularity to appear,
for example when $r=1$ and $\omega_1t=\pi$.

\section{$g\neq0$ case}
\begin{figure}[tbp]
\begin{overpic}[width=4cm]{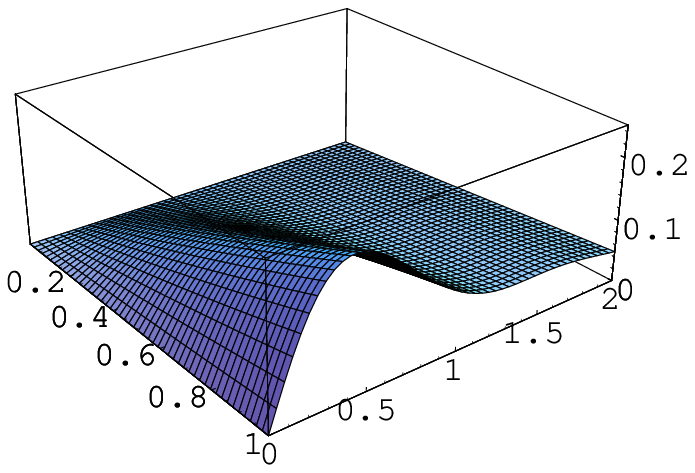}
\put(80,10){$J$} \put(10,15){$r$} 
\put(0,75){(a)}
\end{overpic}
\begin{overpic}[width=4cm]{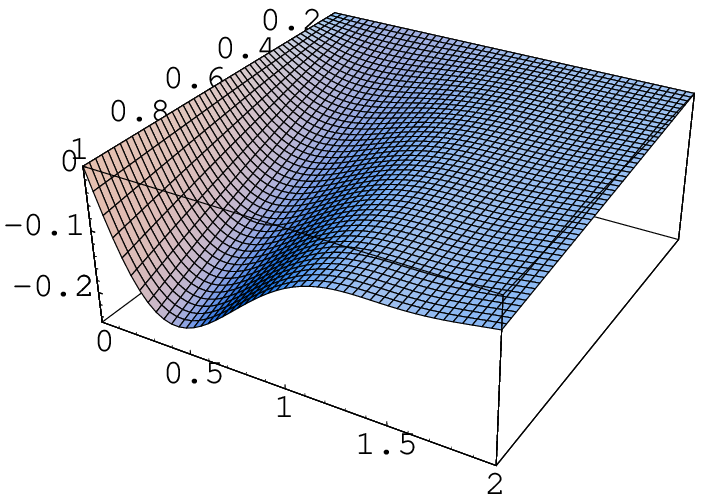}
\put(40,5){$J$} \put(20,70){$r$} \put(-5,60){$\gamma_{g2}$}
\put(0,75){(b)}
\end{overpic}
\caption{\label{3} GP for $\gamma_{g2}$[Arc] versus $r$ and rescaled
coupling constant $J$. We have chosen  $\theta=\pi/4$ (a) and
$\theta=3\pi/4$ (b) for the highlight of the couplings effect on the
geometric phase, which corresponds to the Werner state as the
initial state, and set $\omega_{1}t =2\omega_{2}t =\pi$.}
\end{figure}

Because of the intra-subsystem coupling, the evolution of the system
can be very different from the free case. So in this section we
focus on the effect of coupling on the geometric phase. Obviously
one note that since $[H_I, \rho_1]=0$,  $H_I$ has trivial effect on
$\gamma_{g1}$ and thus this state have been excluded in this
section. Based on the formulas in Sec. II, GP are for $\rho_2$,
\begin{eqnarray}
\gamma_{g2}= \arctan\frac{-
r(\sin\lambda_1t\cos\lambda_2t-\frac{\lambda_2}{\lambda_1}\cos\lambda_1t\sin\lambda_1t)}
{\frac{1-r}{2} +
\frac{1+r}{2}(\cos\lambda_1t\cos\lambda_2t+\frac{\lambda_2}{\lambda_1}\sin\lambda_1t\sin\lambda_1t)}
\end{eqnarray}
in which
\begin{eqnarray}
\lambda_1=&[(\omega_1-\omega_2)\cos2\theta-2g\sin2\theta]/2 \nonumber\\
\lambda_2=&\sqrt{(\omega_1-\omega_2)^2+4g^2}/2.\nonumber
\end{eqnarray}

\begin{figure}
\begin{center}
\begin{overpic}[width=8cm]{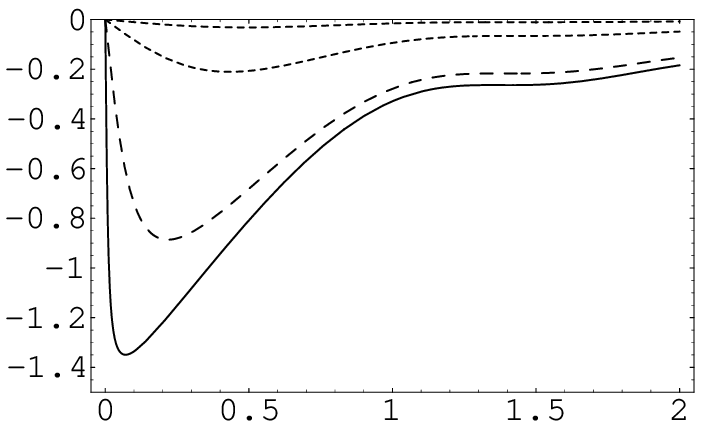}
\put(-15,80){$\gamma_{g2}$} \put(145,-5){$J$}
\put(40,18){n$\rightarrow 0$} \put(40, 50){n=0.1}
\put(55,110){n=0.5} \put(69,125){n=0.8}
\put(100,20){\includegraphics[width=4cm]{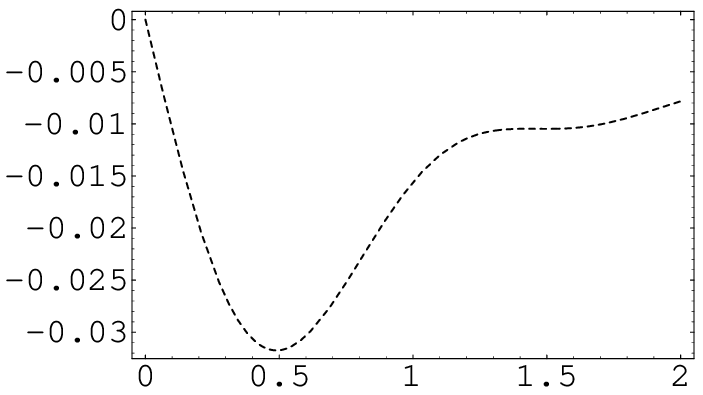}}
\end{overpic}
\caption{\label{4} GP[Arc] vs.  the inhomogeneity of magnetic field
$n=\omega_2/\omega_1$ . We have choose the Werner state
($\theta=\pi/4$)as the initial state with $r=1, \omega_1t=\pi$.The
inset is for $n=0.8$.  Similar behavior can be found when
$\theta=3\pi/4$.}
\end{center}
\end{figure}

The results have been illustrated in Fig. \ref{3} with rescaled
coupling constant $J=\frac{g}{\omega_1}$. Compared with the $g=0$
case, we have chosen the Werner states ($\theta=(3)\pi/4$) as the
initial states and $\omega_1 t=2\omega_2 t=\pi$ with consideration
of the inhomogeneity of external magnetic field. Because of the
coupling, the geometric phase for the Werner state is obviously very
different from the free case. From these figures we also note that
GP has a maximal or mininmal values when the rescaled coupling
constant is less than 1. Furthermore, when $J>1$ the absolute value
of GP decrease with the increment of $J$. One also notes that
because of the mixing, which is scaled by the parameter $r$, the
absolute values of GP is compressed, which is same as the free case.
The effect of the inhomogeneity of magnetic field also discussed.
Our calculation shows that with $\omega_2/\omega_1 \rightarrow 1$,
the point that GP has maximal or minimal value is infinitely closed
to $J=0.5$ (Fig. \ref{4}).

\begin{figure}[tbp]
\begin{overpic}[width=4cm]{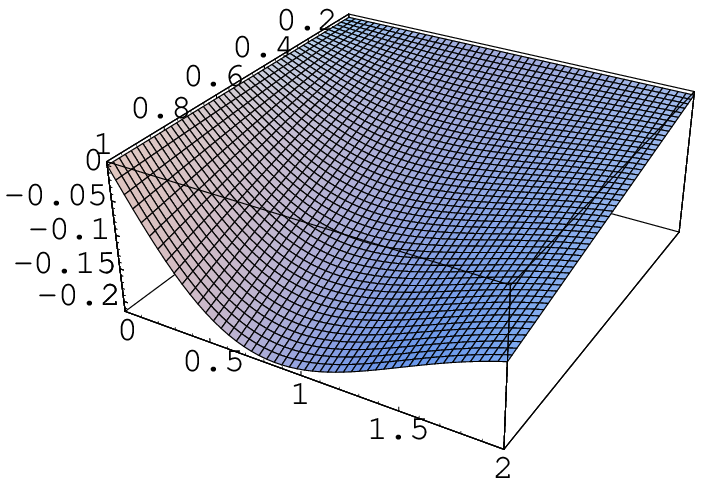}
\put(40,5){$J$} \put(30,75){$r$} \put(0,60){$\gamma_{g2}$}
\put(0,75){(a)}
\end{overpic}
\begin{overpic}[width=4cm]{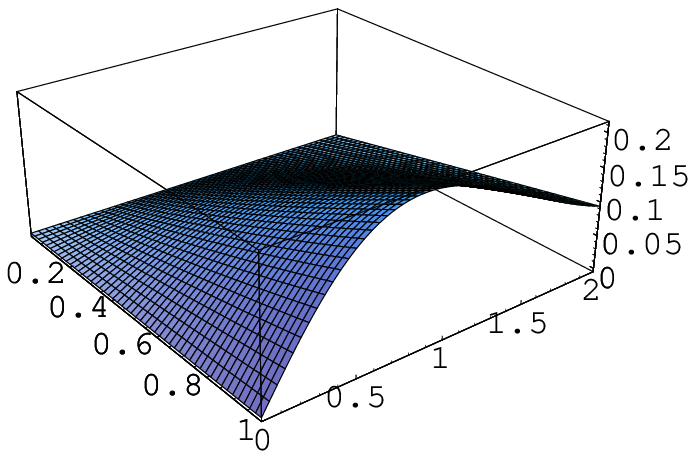}
\put(70,5){$J$} \put(10,15){$r$} \put(100,60){$\gamma_{g2}$}
\put(0,75){(b)}
\end{overpic}
\caption{ \label{5} The geometric phase [Arc]for $\rho_2$ when
$\omega_2=0$ versus  $r, J$. Except of $\omega_2=0$, all other
settings are the same as in Fig. \ref{3}. }
\end{figure}

We also investigate the GP when one particle in zero field. This
case is nontrivial since any system cannot avoid the affect from the
other party. Also it can be used to manipulate the behavior of one
particle by changing of the coupling strength. To this end, we set
$\omega_2 =0$ and the geometric phase have been illustrated in Fig.
\ref{5}. Different from Fig.  \ref{3}, the absolute value of GP
attains the maximal value when $J\rightarrow1$.

\begin{figure}[tbp]
\begin{overpic}[width=4cm]{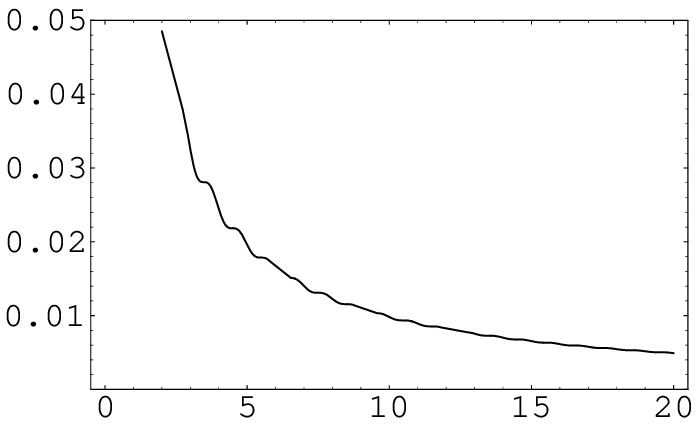}
\put(15, 55){(a)}\put(-5,35){\begin{rotate}{90}$\gamma_{g2}$
\end{rotate}} \put(60, -5){$J$}
\end{overpic}
\hspace{1em}
\begin{overpic}[width=4cm]{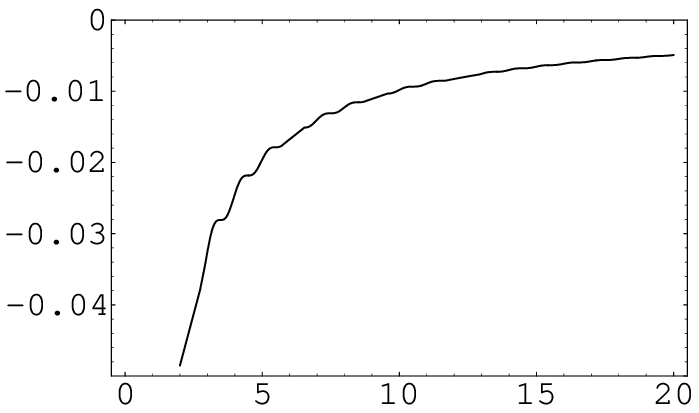}
\put(15,55){(b)}\put(-5,35){\begin{rotate}{90}$\gamma_{g2}$\end{rotate}}
\put(60, -5){$J$}
\end{overpic}
\caption{\label{6} $\gamma_{g2}$[Arc] in the limit of large $J$. The
parameters have same values to that of Fig. \ref{3} and $r=1$ is
chosen.}
\end{figure}

The interesting  extension to this discussion above is in the limit
of large $J=\frac{g}{\omega_1}$. In this case GP has a novel
character, which have been displayed in Fig. \ref{6}.  We note that
GP tends to be zero in the limit of large $J$.  This can be
understood as that in the limit of large  $J$ the interaction $H_I$
is dominant in Eq. (\ref{h})and the Hamiltonian is inclined to be
commutative with the Werner state. Then the entanglement of the
initial state tends  to be unchanged.

Based on the analysis above, we can conclude that because of the
inter-subsystem couplings, the geometric phase shows two different
characters; in the weak coupling limit, the interaction obviously
benefits the geometric phase, displayed in Fig. \ref{3} and Fig.
\ref{5}. However, with further increment  of the couplings, the GP
decreases and tends to be zero infinitely. This phenomena can be
explained easily only if one note that in the infinite coupling
limit, $H_I$ is dominant and its eigenstates are just two of the
Bell states. The entanglement is destroyed in the weak couplings
limit and then revived by the interaction in the infinite limit. The
interesting relation between entanglement and interaction between
two parties is an important aspect of quantum information.

\section{conclusions}
In conclusion, we have discussed the geometric phase for entangled
mixed state Eq. (\ref{rho}) in an  external magnetic field. For the
free case ($g=0$), our studies show that because of the
entanglement, the geometric phase for the system displays two
different types of  character. The first is that if there was no
singularity, the absolute value of the geometric phase attains the
maximum when the initial state is not the Werner state, independent
of the inhomogeneity of the external field (see Fig. \ref{2}(c) and
(d)). Because of the mixing, scaled by $r$, it is compressed.
Furthermore the geometric phase is always zero for the Werner state.
The second behavior occurs when the singularity appears, induced by
the entanglement of the initial states under the proper evolution.
In this case the geometric phase reaches the maximum with $\theta$
tending to the singular point (see Fig. \ref{1}(a) and Fig.
\ref{2}(a) and (b)). We also discuss the situation that one particle
is processing and another keeps dynamically fixed, and similar
results can be found.

For the case $g\neq0$, we choose the Werner states as the initial
condition for facilitating our discussion. The results show that GP
is completely determined by the rescaled coupling constant $J$ and
attain the maximal or minimal value when $J<1$. Similar to the free
case, we also discussed the effect of the inhomogeneity of the
external magnetic field. The results show that for Werner state GP
is always zero when  system is in  a homogeneous external magnetic
field , independent of the interaction. Furthermore with
$\omega_2/\omega_1\rightarrow1$, the absolute value of GP attains a
maximum at the point $J\rightarrow0.5$. The geometric phase for
$\omega_2=0$ is also studied. Further study in the limit of large
$J$ displays a novel phenomena that GP tends to be zero and this
result have been explained as the revival of entanglement in this
limit.

This work was supported by NSF of China under grant 10305002 and
60578014.

\end{document}